\documentclass[%
 reprint,
 superscriptaddress,
 bibnotes,
 amsmath,amssymb,
 aps,
 citeautoscript,
 floatfix,
]{revtex4-2}

\usepackage{graphicx}
\usepackage{dcolumn}
\usepackage{bm}
\usepackage[dvipsnames]{xcolor}
\usepackage[colorlinks=true, citecolor={blue!80!black}, urlcolor={blue!50!black}, linkcolor = {blue!80!black}]{hyperref}
\usepackage{lipsum}  
\usepackage{upgreek}
\usepackage[separate-uncertainty=true]{siunitx}
\usepackage{mathtools}
\usepackage{ulem}
\usepackage{braket}

\usepackage{nameref,zref-xr}
\zxrsetup{toltxlabel}

\newcommand{\abs}[1]{\left|#1\right|}
\newcommand{\Vb}{$V_{\rm bias}$}
\newcommand{\Qna}{$\rm{Near}_{\rm Al}$}
\newcommand{\Qnn}{$\rm{Near}_{\rm NbTiN}$}
\newcommand{\Qfn}{$\rm{Far}_{\rm NbTiN}$}
\newcommand{\Qfa}{$\rm{Far}_{\rm Al}$}

\makeatletter
\newcommand*{\addFileDependency}[1]{
  \typeout{(#1)}
  \@addtofilelist{#1}
  \IfFileExists{#1}{}{\typeout{No file #1.}}
}
\makeatother

\newcommand*{\myexternaldocument}[1]{%
    \zexternaldocument*{#1}%
    \addFileDependency{#1.tex}%
    \addFileDependency{#1.aux}%
}

\begin{document}

\myexternaldocument{supplement}
\preprint{APS/123-QED}

\title{Mitigation of quasiparticle loss in superconducting qubits by phonon scattering}

\author{Arno Bargerbos}
 \email{a.bargerbos@tudelft.nl}
\affiliation{QuTech and Kavli Institute of Nanoscience, Delft University of Technology, 2628 CJ Delft, The Netherlands}
\author{Lukas Johannes Splitthoff}
\affiliation{QuTech and Kavli Institute of Nanoscience, Delft University of Technology, 2628 CJ Delft, The Netherlands}
\author{Marta Pita-Vidal}
\affiliation{QuTech and Kavli Institute of Nanoscience, Delft University of Technology, 2628 CJ Delft, The Netherlands}
\author{Jaap J. Wesdorp}
\affiliation{QuTech and Kavli Institute of Nanoscience, Delft University of Technology, 2628 CJ Delft, The Netherlands}
\author{Yu Liu}
\affiliation{Center for Quantum Devices, Niels Bohr Institute, University of Copenhagen, 2100 Copenhagen, Denmark\looseness=-1}
\author{Peter Krogstrup}
\affiliation{Niels Bohr Institute, University of Copenhagen, 2100 Copenhagen, Denmark\looseness=-1}
\author{Leo P. Kouwenhoven}
\affiliation{QuTech and Kavli Institute of Nanoscience, Delft University of Technology, 2628 CJ Delft, The Netherlands}
\author{Christian Kraglund Andersen}
\affiliation{QuTech and Kavli Institute of Nanoscience, Delft University of Technology, 2628 CJ Delft, The Netherlands}
\author{Lukas Gr\"unhaupt}
 \altaffiliation[Now at: ]{Physikalisch-Technische Bundesanstalt, 38116 Braunschweig, Germany}
\email{lukas.gruenhaupt@ptb.de}
\affiliation{QuTech and Kavli Institute of Nanoscience, Delft University of Technology, 2628 CJ Delft, The Netherlands}

\date{\today}

\begin{abstract}
Quantum error correction will be an essential ingredient in realizing fault-tolerant quantum computing. However, most correction schemes rely on the assumption that errors are sufficiently uncorrelated in space and time. In superconducting qubits this assumption is drastically violated in the presence of ionizing radiation, which creates bursts of high energy phonons in the substrate. These phonons can break Cooper-pairs in the superconductor and, thus, create quasiparticles over large areas, consequently reducing qubit coherence across the quantum device in a correlated fashion. A potential mitigation technique is to place large volumes of normal or superconducting metal on the device, capable of reducing the phonon energy to below the superconducting gap of the qubits. To investigate the effectiveness of this method we fabricate a quantum device with four nominally identical nanowire-based transmon qubits. On the device, half of the niobium-titanium-nitride ground plane is replaced with aluminum (Al), which has a significantly lower superconducting gap. We deterministically inject high energy phonons into the substrate by voltage biasing a galvanically isolated Josephson junction. In the presence of the low gap material, we find a factor of 2-5 less degradation in the injection-dependent qubit lifetimes, and observe that undesired excited qubit state population is mitigated by a similar factor. We furthermore turn the Al normal with a magnetic field, finding no change in the phonon-protection. This suggests that the efficacy of the protection in our device is not limited by the size of the superconducting gap in the Al ground plane.
Our results provide a promising foundation for protecting superconducting qubit processors against correlated errors from ionizing radiation.
\end{abstract}

\maketitle
Superconducting qubits are one of the prime candidates in the global effort towards building a quantum computer. Tremendous technological advances have been achieved over the last decade, heralding the advent of noisy intermediate-scale quantum technologies \cite{Preskill2018, Arute2019, Kjaergaard2020}. In order to go beyond this intermediate scale and harness the full potential of quantum computers, fault-tolerant quantum computing will be required. Remarkable progress has been made in terms of implementing error detection and correction in recent years using superconducting circuits \cite{Andersen2020, Chen2021, Marques2021, Krinner2022}.
A key assumption of most quantum error correction schemes is that qubit errors are spatially and temporally uncorrelated, however that appears to be drastically violated in large scale superconducting qubit arrays. In \cite{McEwen2021} it has been shown that cosmic rays and ambient radioactivity can deposit large amounts of energy into the substrate of the device in the form of phonons. These phonons travel over distances of centimeters, breaking up Cooper pairs and leading to decreased qubit coherence over timescales of milliseconds, causing correlated error events \cite{Swenson2010, Moore2012, Vepsaelaeinen2020, McEwen2021, Wilen2021, Martinis2021, Cardani2021}.

Several mitigation strategies have been proposed to combat these ionizing impact events at the level of the quantum device, such as the direct trapping of quasiparticles through gap engineering \cite{Wang2014, Riwar2016, Riwar2019, Pan2022} as well as impeding the propagation of phonons by substrate modification \cite{Chu2016, Rostem2018, Karatsu2019, Puurtinen2020}. A complementary approach is the use of so-called phonon traps \cite{Patel2017, Henriques2019, Karatsu2019, Martinis2021, Visser2021}. Made from a normal or superconducting material with a small superconducting gap, phonon traps dissipate the phonon energy through scattering events until the resulting phonons have too little energy to break Cooper pairs in the qubit layer. A key difference of phonon traps compared to the direct trapping of quasiparticles is that the traps target the phonons also while they are en-route to the qubits, before error events occur. Phonon traps furthermore do not have to be galvancially connected to the circuit, nor do they even have to be embedded into the same plane of the chip, as long as they connect to the substrate. It is therefore possible to design phonon traps without introducing qubit dissipation from coupling to lossy materials \cite{Riwar2019}, and with no added strain on the increasing complex task of control line routing \cite{Brecht2016}. 

\begin{figure}[t!]
\centering
\includegraphics[scale=1.0]{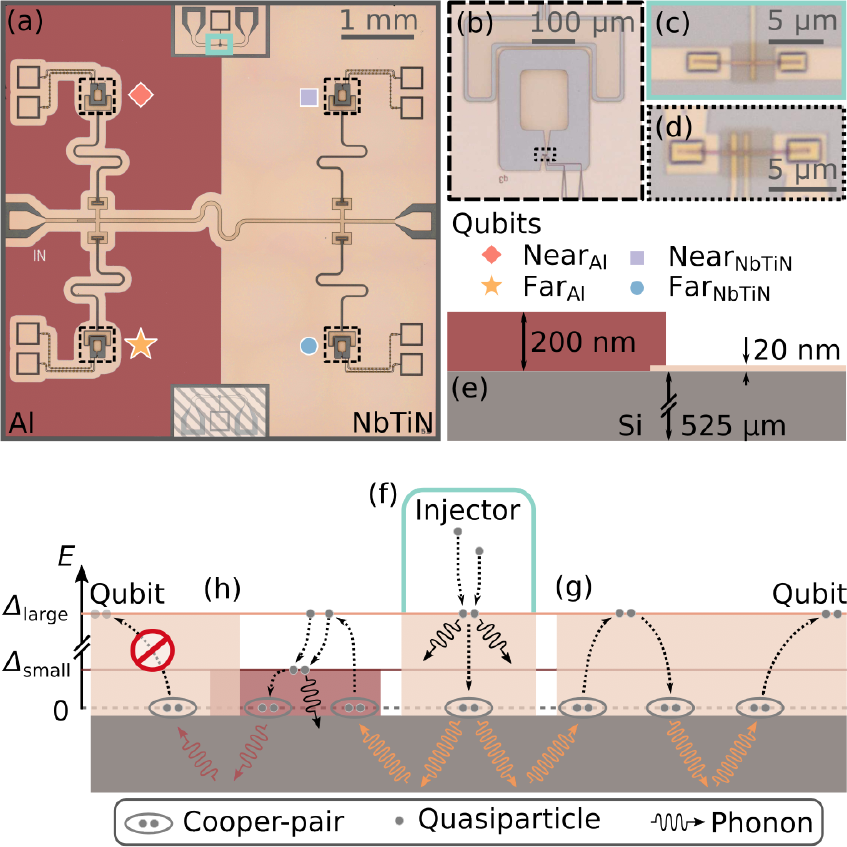}
\caption{\label{fig:device} False colored device overview (a-e) and sketch of the phonon mitigation process (f-h). (a) Four nanowire transmon qubits [dashed boxes, see (b)] are coupled to individual readout resonators which in turn are coupled to a common feedline. Two nanowire-based junctions are additionally present for phonon injection [turquoise boxes, see (c)] and are galvanically isolated from the qubit ground plane (bottom junction not operational). All structures are patterned from NbTiN (light orange), except for the majority of the left half of the ground plane, which is made from a thick aluminum film (red) [see device cross-section in (e), lateral dimension not to scale]. Electrostatic gates below the junctions tune the resistance of the phonon injector (c) and Josephson energy of the qubits [d, zoom-in of (b)], respectively. A second gate per qubit allows for independent tuning of the qubit charge offsets. 
(f) Quasiparticles excited by voltage biasing the injector relax to the superconducting gap edge and recombine, emitting phonons (black and orange arrows, respectively). (g) The emitted phonons propagate via the substrate, and those with energies larger than twice the superconducting gap $> 2\Delta_\mathrm{large}$ induce Cooper-pair breaking, relaxation and recombination cycles, eventually exciting quasiparticles in the qubits. (h) In the presence of a small gap superconductor, these cycles produce phonons of energy $\leq 2\Delta_\mathrm{small}$ instead (red arrows), which cannot break Cooper-pairs in the qubits.}
\end{figure}

To date, the efficacy of phonon traps has been demonstrated for superconducting resonators \cite{Patel2017, Henriques2019} and kinetic inductance detectors \cite{Karatsu2019}. In this article we set out to investigate their effectiveness in protecting superconducting qubits by fabricating a $6 \times 6~\rm{mm^2}$ chip containing four nanowire transmon qubits \cite{deLange2015, Larsen2015}, one in each corner of the device [Fig.~\ref{fig:device}(a)]. All transmons have identical geometries [Fig.~\ref{fig:device}(b)] and are excited and read out via individual coplanar waveguide resonators coupled to a common feedline. The transmon islands, resonators and feedline are dry etched from a \SI{20}{\nano\meter} thick niobium-titanium-nitride (NbTiN) thin film deposited on a \SI{525}{\micro\meter} thick high-resistivity silicon substrate. We implement the phonon traps by partially removing the NbTiN groundplane on one half of the chip, replacing it with \SI{200}{\nano\meter} of aluminium (Al) deposited by electron beam evaporation and patterned by lift-off [cf. Fig.~\ref{fig:device}(e)]. The Al and NbTiN layers are furthermore galvanically connected by a \SI{20}{\micro\meter} wide region of overlap between them. We leave a \SI{290}{\micro \meter} region of NbTiN around all qubit islands to suppress direct quasiparticle trapping in the Al layer surrounding the qubit \cite{Riwar2016, Riwar2019}. While quasiparticle trapping can be of great use in practical qubit applications, the goal of this study is to evaluate the effect of phonon trapping only. Finally, the full backside of the chip is glued to a solid copper block using thermally conductive silver epoxy. Through the copper, the chip and surrounding enclosure are thermally anchored to the mixing chamber of a dilution refrigerator at $\sim \SI{20}{mK}$.

Each transmon island is connected to ground via a nominally \SI{10}{\micro m} long epitaxial semiconductor-superconductor nanowire, consisting of a \SI{110}{nm} wide hexagonal InAs core and a \SI{6}{nm}-thick Al shell covering two of its facets \cite{Krogstrup2015}. By selectively removing a \SI{100}{\nano\meter} long segment from the Al shell we define a semiconducting Josephson junction, whose Josephson energy can be tuned with a single bottom gate electrode via the field effect \cite{deLange2015, Larsen2015}. A second gate electrode is present under an InAs-Al region of the nanowire, allowing for capacitive tuning of the island's offset charge and aiding in the estimation of the qubit parameters \cite{Note3}. The choice for nanowire-based junctions over conventional tunnel junctions is motivated by their magnetic field compatibility \cite{Luthi2018, PitaVidal2020, Uilhoorn2021}, allowing us to study the dependence of phonon trapping efficacy on the size of the superconducting gap in the \SI{200}{nm} thick Al ground plane without strongly affecting the qubit parameters \cite{Note3}. 

A key feature of our device is two additional semiconducting junctions used for phonon injection. Identical to those used for the transmons, the junctions are located at the top and bottom of the front side of the chip, but galvanically isolated from the qubit ground plane [see Fig.~\ref{fig:device}(a)]. They are connected to source and drain leads made from the NbTiN base-layer to allow for voltage biasing and current sensing, while the resistance of the junctions can be tuned with a bottom gate electrode \cite{Note3}. In this article only the top junction is used, as the bottom junction is not functional. The top junction serves to inject highly energetic phonons into the substrate of the chip, a technique pioneered in the context of non-equilibrium phonon dynamics \cite{Eisenmenger1967, Welte1972, Eisenmenger1976} that has more recently been adapted for experiments involving superconducting quantum circuits \cite{Lenander2011, Wenner2013, Patel2017}. Given that the four qubits can be uniquely identified based on their distance from the top junction (\SI{1.8}{\milli\meter} or \SI{4.4}{\milli\meter}) and whether their ground plane is made from Al or NbTiN, we label them as the \Qna, \Qnn, \Qfa, and \Qfn~qubits \cite{Note3}.

\begin{figure}[t!]
\centering
\includegraphics[scale=1.0]{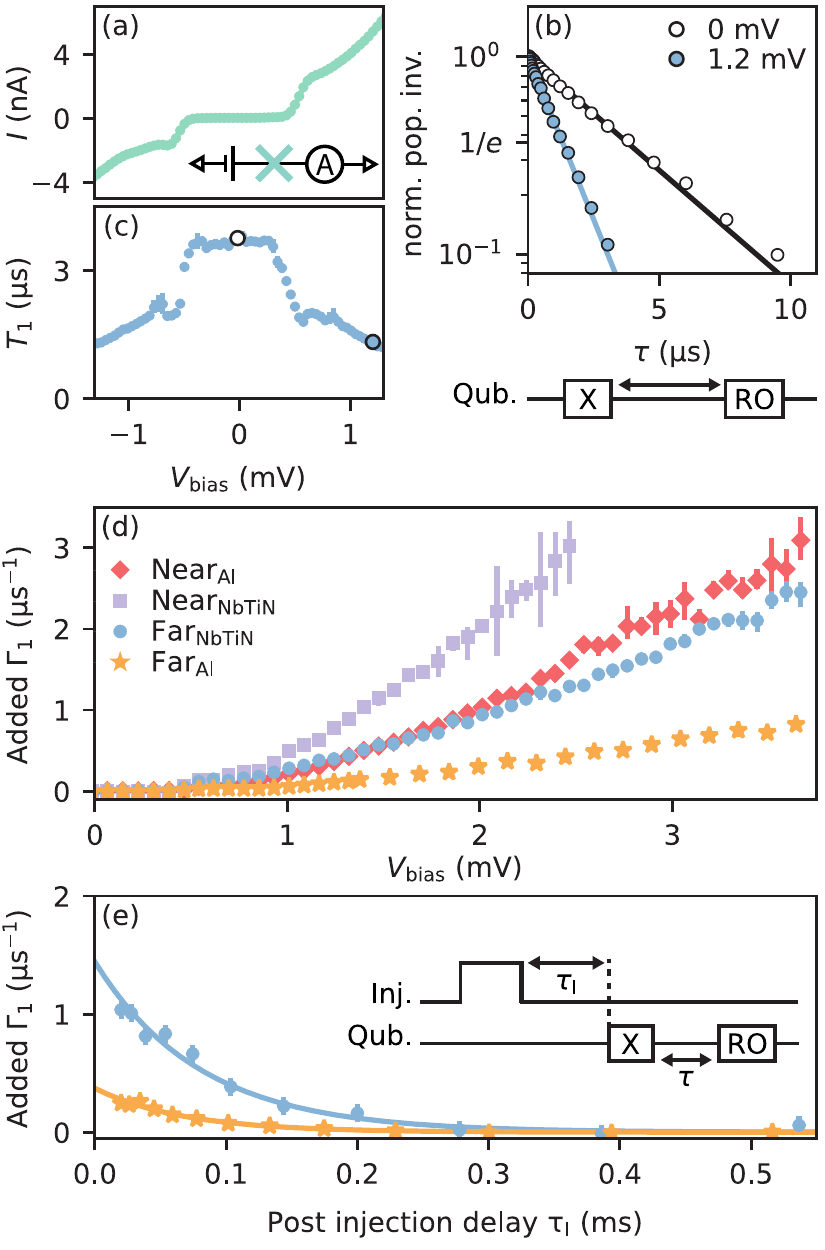}
\caption{\label{fig:lossrates_DC} Added qubit loss due to phonon injection. (a) I-V characteristics of the Josephson junction used to inject phonons measured using the circuit shown in the inset. (b) Representative qubit lifetime ($T_1$) measurements on the $\mathrm{Far}_{\rm{NbTiN}}$ qubit for bias voltages \Vb~of 0 and \SI{1.2}{\milli\volt} applied across the injector. The data (markers) is fit with an exponential decay (solid lines), yielding lifetimes of \SI{3.8}{\micro s} and \SI{1.3}{\micro s} respectively. (c)  Qubit lifetime $T_{1}$ of the $\mathrm{Far}_{\rm{NbTiN}}$ qubit as a function of \Vb. (d) Added qubit loss rate $\Gamma_{1}$ as a function of \Vb~for all 4 qubits (labels apply to all panels). We define added $\Gamma_{1}$ as the bias voltage dependent loss rate minus the average base line value for $|eV_\mathrm{bias}| < 2\Delta_{\rm nw}$ [cf. panel (c)]. (e) Added $\Gamma_{1}$ as a function of the delay time $\tau_{\rm{I}}$ after a square injection pulse with a duration of \SI{20}{\micro s} and an amplitude of $\approx \SI{3}{mV}$. The loss rates are fit with an exponential decay (solid line), yielding recovery times of $\SI{80\pm9}{\micro s}$ (blue) and $\SI{67\pm5}{\micro s}$ (orange). Error bars in panels (c-e) denote the standard deviation over 5 repetitions.}
\end{figure}

As the phonon injection and propagation process is central to this experiment, we briefly discuss the operation mechanism here, which is visually represented in Fig.~\ref{fig:device}(f-h). We set the junction's bottom gate voltage such that we operate in the tunneling regime, where the supercurrent at bias voltage $V_{\rm bias}=0$ is fully suppressed \cite{Note3}. For voltages $V_{\rm bias} < \abs{2\Delta_{\rm nw}}/e$ applied between the leads of the junction there is thus no current, where $\Delta_{\rm nw}$ is the superconducting gap of the InAs-Al nanowires. However, for voltages $V_{\rm bias} \geq \abs{2\Delta_{\rm nw}}/e$ Cooper pairs can be broken up and a quasiparticle current will run across the junction [see Fig.~\ref{fig:lossrates_DC}(a)]. The quasiparticles then diffuse around the vicinity of the junction and its leads, scattering and relaxing to the gap edge of the superconductors, producing relaxation phonons of energies up to $E=eV_{\rm bias} - \Delta_{\rm nw}$ in the process, see Fig.~\ref{fig:device}(f) \cite{Otelaja2013}. The now-relaxed quasiparticles can subsequently also recombine with other quasiparticles, emitting recombination phonons of energy $E=2\Delta_{\rm nw}$ \footnote{When quasiparticles recombine they can also emit photons rather than phonons. However, as the final density of states corresponding to this process is much smaller than the density of states for phonon emission, we neglect this effect.}. These phonons either break up new Cooper pairs in the metal layer, or they escape into the substrate where they can rapidly travel over distances of several times the size of the chip, scattering off the boundaries \cite{Martinez2019}. The phonons can thus end up at the qubits, creating quasiparticles and inducing losses proportional to the excess quasiparticle density $x_{\rm qp}$ \cite{Glazman2021}, see Fig.~\ref{fig:device}(h). However, if the phonons encounter the Al traps en-route to the qubits, part of their energy can be dissipated in further cycles of Cooper pair breaking, relaxation, and recombination [Fig.~\ref{fig:device}(g)]. The resulting phonons of energy $E \leq 2\Delta_{\rm trap}$ can no longer excite quasiparticles in the qubit structures, for which the superconducting gaps of the islands $\Delta_{\rm NbTiN}\geq \SI{1500}{\micro eV}$ and of the nanowires $\Delta_{\rm nw} = \SI{270}{\micro eV}$ are larger than that of the traps $\Delta_{\rm trap} = \SI{180}{\micro eV}$ \cite{Woerkom2015, Marchegiani2022, Splitthoff2022, Note3}. 

To investigate the effectiveness of the Al ground plane to protect against high energy phonons, we perform qubit lifetime ($T_1$) experiments on all four qubits while we apply a constant bias to the phonon injector, which causes the $T_1$ time to decrease as shown Fig.~\ref{fig:lossrates_DC}(b). For bias voltages within the gap, $T_1$ remains large constant, see Fig.~\ref{fig:lossrates_DC}(c), while $T_1$ drastically decreases at the onset of the quasiparticle current, with a kink at $2\Delta_{\rm nw}$ originating from the enhanced conductance at the gap edge \cite{Note3}. As the qubit is over \SI{4}{mm} away from the galvanically isolated injector junction, this suggests that the losses indeed originate from phonons that traveled through the substrate \cite{Patel2017}. We then compare the added loss rate $\Gamma_1(V_{\rm bias}) = 1/T_1(V_{\rm bias}) - 1/T_{1}(V_{\rm bias} < \vert 2\Delta_{\rm nw}/e \vert)$ of each qubit, where we subtract the baseline loss rate measured inside the superconducting gap. As can be seen in Fig.~\ref{fig:lossrates_DC}(d), the \Qfa~qubit has up to 8 times smaller added loss rate than the \Qnn~qubit. However, this comparison involves both a larger separation from the injector and the presence of phonon traps. To disentangle the two effects, we compare the qubits at equal distance from the injector. The \Qna~qubit has up to 2.5 times smaller added loss rate than the \Qnn~qubit, supporting that the presence of the Al phonon traps leads to resilience against phonon induced losses. For the \Qfn~and \Qfa~qubits the improvement even reaches up to 5, suggesting that an increased area of the trapping region increases the efficacy \cite{Note3}. We note that the improvements are bias dependent, tending towards a constant value for bias voltages above \SI{1.5}{\mV}, several times the size of $\Delta_{\rm nw}$.

So far, we have focused on phonons that are continuously injected into the chip by applying a constant bias voltage across the injector. However, in the impact events of ionizing radiation the phonons are created in bursts. To test whether our findings still hold under such circumstances, we repeat the same experiment using a pulsed phonon injection scheme. Now we apply a square pulse of duration $\SI{20}{\micro s}$ and an amplitude of $\approx \SI{3}{mV}$ across the injector, and we subsequently measure the qubit loss as a function of delay time after the injection event [Fig.~\ref{fig:lossrates_DC}(e)]. The recovery of added loss rate with delay time follows an exponential form, suggesting a recovery dominated by quasiparticle relaxation rather than by recombination, for which the recovery is governed by a hyperbolic cotangent function \cite{Wang2014}. We find that at zero delay time the \Qfn~qubit is affected 4 times more than the \Qfa~qubit, consistent with the results for continuous injection. We further note that that the recovery times are approximately equal at $\SI{80\pm9}{\micro s}$ [$\SI{67 \pm 5}{\micro s}$] for the \Qfn~[\Qfa] qubit. We do not observe any significant enhancement of recovery time due to the presence of the Al traps, such that the recovery time might instead be dominated by quasiparticle trapping or by the time it takes for phonons to leave the substrate. Alternatively, the phonon traps could have a similar effect on recovery times across the device, as the phonons can traverse the chip on timescales of a few microseconds \cite{Martinis2021}. 

\begin{figure}[t!]
\centering
\includegraphics[scale=1.0]{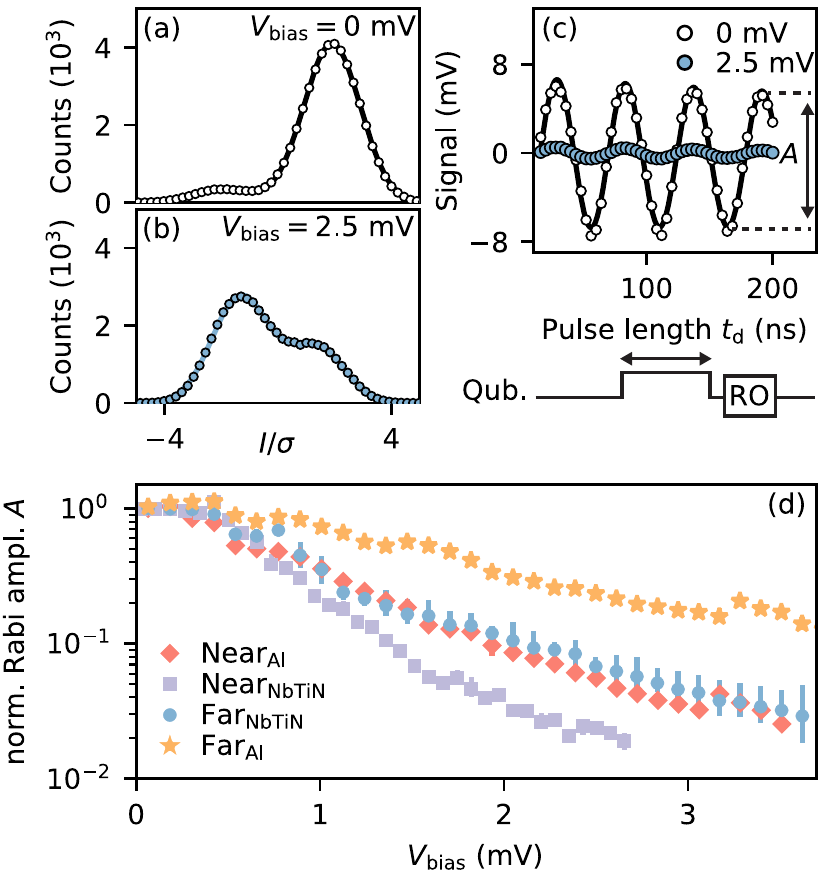}
\caption{\label{fig:rabi} Increased excited state population due to phonon injection. (a) Histogram of the resonator response for the $\mathrm{Far}_{\rm{NbTiN}}$ qubit at $V_\mathrm{bias} = \SI{0}{mV}$. Each individual shot corresponds to the integrated output signal for a readout time of \SI{500}{ns}. The resulting histogram (markers) is fit with a double Gaussian function (solid line), from which we estimate a ground state population of 93\%. (b) Same as (a) for $V_{\rm bias}=\SI{2.5}{mV}$, from which we estimate a ground state population of 35\%. (c) Representative time-Rabi experiments on the $\mathrm{Far}_{\rm{NbTiN}}$ qubit for the values of \Vb~used in panels (a-b). The data (markers) is fit with an exponentially decaying cosine (solid line). (d) Extracted Rabi oscillation amplitude $A$ normalized to its value at zero bias, measured  as a function of \Vb~for all 4 qubits. Error bars in panel (d) denote the standard deviation over 5 repetitions.}
\end{figure}

In the measurements of the qubit loss rates we additionally find that the qubit readout signal becomes smaller with increasing bias voltages, requiring substantially more repetitions to obtain the same signal-to-noise ratio (SNR) at elevated bias. To investigate this effect we monitor the resonator response of the \Qfn~qubit using single-shot readout in the absence of any qubit excitation tones. For \SI{0}{mV} bias this results in a double Gaussian distribution of measurement outcomes, with 93\% of the outcomes located in a single Gaussian [Fig~\ref{fig:rabi}(a)]. We interpret this output as the signal corresponding to the ground state population of the transmon qubit, indicating a residual excited state population of 7\%. When we apply a constant bias of \SI{2.5}{mV} the distribution changes significantly: the transmon is now in the ground state only 35\% of the time. This is consistent with previous findings showing that energetic non-equilibrium quasiparticles can lead to an increased excited state population \cite{Wenner2013, Jin2015, Serniak2018}. The fact that the ground state population is less than 50\%  could furthermore indicate that the transmon now also has a sizeable population outside of the two-level qubit subspace, and that high energy phonons thus also cause qubit leakage errors, which are not easily mitigated in standard error correction schemes \cite{Varbanov2020, McEwen2021a}. However, the SNR of our measurements is not large enough to distinguish the different excited transmon states, in particular for elevated bias voltages that strongly reduce the qubit lifetime as well as potentially populating a plethora of states. 

To gain insight into the effectiveness of the traps in reducing unwanted excited population, while constrained by limited SNR, we perform a time-Rabi experiment between the transmons' ground and first excited state as a function of the injection bias voltage. The amplitude of the Rabi oscillation $A$ is proportional to the difference in the population of the transmon states involved, and additionally decreases in the presence of qubit leakage. The evolution of $A$ with $V_{\rm bias}$ is thus indicative of the change in the state populations. As shown in Fig~\ref{fig:rabi}(c), the amplitude of the oscillations for the \Qfn~qubit indeed decreases significantly at $V_{\rm bias} = \SI{2.5}{mV}$ compared to that at \SI{0}{mV}. We repeat this experiment for all qubits, normalizing the Rabi amplitude to its value at zero injection voltage. We find results comparable to those of the added qubit loss rates, with the \Qna~qubit again showing a reduction in Rabi amplitude of up to 5 times larger than that of the \Qfn~qubit. This difference indicates that phonon traps can thus also protect against phonon-induced excited state population. 

\begin{figure}[t!]
\centering
\includegraphics[scale=1.0]{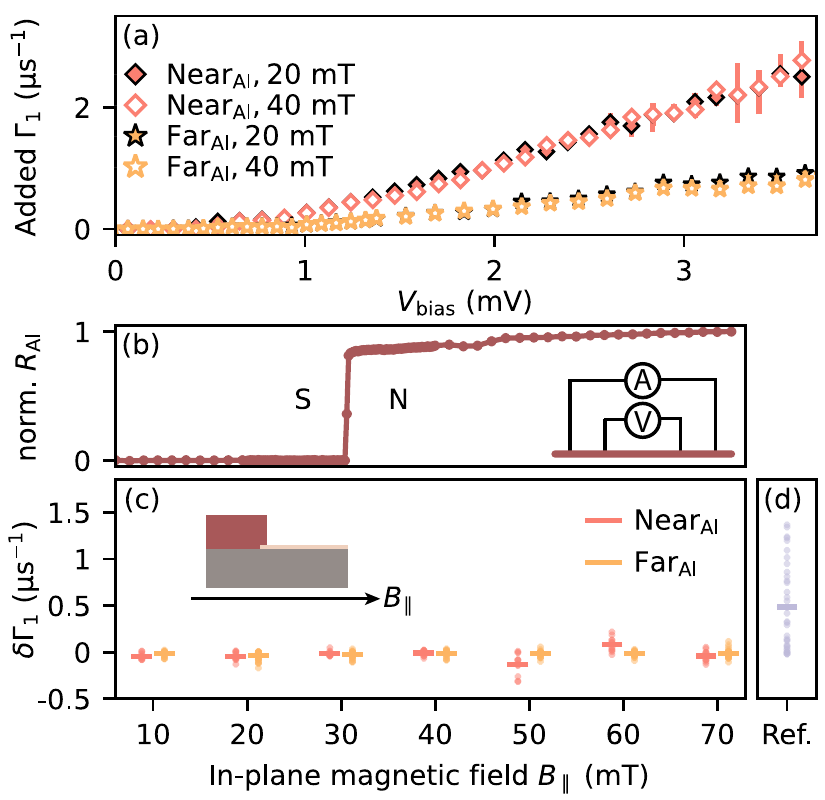}
\caption{\label{fig:field} Magnetic field independence of added qubit loss. (a) Added $\Gamma_{1}$ as a function of \Vb~for the two qubits with an aluminum ground plane at magnetic fields $B_{\rm{\parallel}}$ of 20 and \SI{40}{mT} applied in the plane of the chip [cf. inset panel (c)]. (b) Four-point measurement of the normalized aluminum film resistance $R_{\rm{Al}}$ as a function of $B_{\rm{\parallel}}$. (c) Scatter plot showing the differences between the \Vb~dependence of $\Gamma_{1}$ at 0 and elevated $B_{\rm{\parallel}}$ for the two qubits with an aluminum ground plane. Each data point represents the difference in rates $\delta \Gamma_{1}$ evaluated at equal \Vb. The horizontal line indicates the mean of $\delta \Gamma_{1}$ over all \Vb. Inset: direction of the applied magnetic field with respect to the cross-section of the chip. (d) Same type of plot as (c), but showing the difference between the $\rm{Near}_{\rm Al}$ and $\rm{Near}_{\rm NbTiN}$ qubits at 0 applied magnetic field [data shown in Fig.~\ref{fig:lossrates_DC}(d)]. This serves as a reference for the data in panel (c).}
\end{figure}

Having established a moderate protection due to the Al traps, we investigate how the difference in superconducting gap between the trapping and qubit material influences the effectiveness of the trapping process. For the trapping process to be of use, the superconducting gap of the trap must be smaller than that of the qubit layer [cf. Fig.~\ref{fig:lossrates_DC}(f-h)]. As the relaxation rate of quasiparticles excited inside the trap grows with their energy above the gap edge \cite{Kaplan1976, Riwar2019}, we assume that the absolute size of gap should be relevant. Ignoring potentially detrimental effects from electromagnetic coupling, a normal metal (with no spectral gap) might thus be particularly suitable as a trapping material \cite{Patel2017, Martinis2021}. We investigate this hypothesis in-situ, making use of the inherent magnetic field compatibility of the nanowire transmon qubits and their readout circuit, which have been shown to be operable up to parallel fields in excess of \SI{1}{T} \cite{Luthi2018, Kroll2019, PitaVidal2020, Uilhoorn2021}. This is in contrast to the \SI{200}{\nano\meter} thick Al, which turns normal at significantly lower fields of $\approx \SI{30}{\milli\tesla}$ [see Fig.~\ref{fig:field}(b)] \cite{Meservey1971}.

We repeat the measurements of added loss rate in the presence of a constant bias voltage for different applied magnetic fields. Contrary to expectation, the added loss rate does not appear to change after passing the critical parallel field of the thick Al [Fig.~\ref{fig:field}(a)]. In fact, we do not detect any significant change as a function of magnetic field between 0 and \SI{70}{mT}, far in excess of the critical field [Fig.~\ref{fig:field}(c)]. These findings indicate that the size of the superconducting gap is not what sets the effectiveness of our phonon traps. A possible explanation is that the energy gap of the qubit island $\Delta_{\rm NbTiN}$ is approximately ten times larger than the gap of the Al traps $\Delta_{\rm trap}$, such that the relaxation rate of quasiparticles - and thus the rate at which the traps reduce  phonon energy - already approaches that of the normal state even at zero field \cite{Riwar2019}. However, as shown in Fig.~\ref{fig:lossrates_DC}(c), the lifetime of the qubits is reduced already when $V_{\rm bias} > \abs{2\Delta_{\rm nw}}$, significantly below $2\Delta_{\rm NbTiN}$, for which no quasiparticles can be excited in the NbTiN. This indicates that the phonons excite quasiparticles directly in the proximitized nanowires, for which the quasiparticle relaxation rate should still show an improvement as a function of magnetic field \cite{Martinis2021}. 

We instead hypothesize that the rate at which excited quasiparticles relax inside the phonon trap is not the limiting process, but that the bottleneck for the trapping efficacy is the rate at which phonons create quasiparticles in the trap to begin with. This could be due to the relatively long interaction length of the phonons inside the metal, in which case a significantly thicker trapping layer would be beneficial \cite{Martinis2021}. Alternatively, the trapping could be limited by the poor interface between the substrate and the Al layer \cite{Kaplan1979, Martinez2019}, making it difficult for phonons to enter the traps. In our device this is likely exacerbated by the dry etching procedure used to remove the NbTiN layer before Al deposition, which roughens the underlying silicon, potentially reducing phonon transmission across the metal-substrate interface.

In conclusion, we find a factor of 2-5 improvement in the protection against phonon-induced degradation of qubit lifetimes, as well as in the increase in excited state population for transmon qubits surrounded by Al phonon traps. This level of improvement is in line with previous results on phonon trapping for superconducting resonators \cite{Patel2017, Henriques2019} and kinetic inductance detectors \cite{Karatsu2019}, here demonstrated at the level of sensitivity of transmon qubits. While the obtained improvement is modest, we emphasize that this is a conservative estimate for realistic multi-qubit arrays. With mean-free paths exceeding several times the size of the chip, phonons in the silicon are able to travel vast distances, and thus it is likely that the two NbTiN-based qubits also benefit from the presence of the Al traps. Therefore, we believe the improvement found is a lower bound on what could be obtained when comparing different chips with and without traps, which we choose not to do to exclude unintended differences between devices and thermal anchoring of the respective chips. Furthermore, we deliberately thermally anchor the full backside of the chip, allowing phonons to leave the device via the substrate. This is in contrast to most superconducting qubit implementations, where the chips are mounted in a floating configuration \cite{Wenner2011, Lienhard2019, Martinis2021}. In these devices the main path for the phonons to escape the device is through the wirebonds at the perimeter of the chip, a slow process that enhances the probability for the phonons to interact with the traps in such devices. 

During the writing of this manuscript we became aware of a similar experiment by Iaia et al. \cite{Iaia2022}. In their work the authors compare the impact of high energy phonons on two separate transmon chips, where one chip has a \SI{10}{\micro m} thick Cu film deposited on its backside. For the Cu-covered device they find a reduction in phonon-induced qubit errors by more than a factor of 20. These findings strongly complement the results of this paper, where we instead investigate the effect of superconducting traps. Furthermore, by comparing transmons on a single chip we provide a conservative measure of what trapping efficacy can be achieved. Together these works show that phonon traps offer a promising path towards reducing correlated errors to below the level required for fault-tolerant operation.

Finally, we highlight that phonon traps are also relevant for transmons realized with conventional Al/AlOx Josephson junctions \cite{Marchegiani2022}, as well as for other types of superconducting qubits, such as fluxoniums \cite{Pop2014} and novel protected qubit designs \cite{Gyenis2021}.
The enhanced rates of quasiparticle poisoning events following an impact event are also expected to be highly detrimental for parity-based qubits such as Andreev qubits \cite{Janvier2015, Hays2021}, as well as topologically protected Majorana qubits \cite{Rainis2012, Karzig2021}. The current generation of devices used in these qubit platforms rely on the same type of superconductor-semiconductor nanowires as used in this experiment.
Furthermore, while not directly sensitive to superconductor-based quasiparticles, spin qubits are known to suffer phonon-mediated back-action \cite{Schinner2009, Granger2012} and might also suffer from correlated errors due to phonon impacts, although to what extent remains to be investigated. 

\begin{acknowledgments}
We gratefully acknowledge Olaf Benningshof, Jason Mensingh, and Raymond Schouten for technical support. We further thank Grzegorz Mazur for help with the critical field measurements, Gianluigi Catelani for discussions about quasiparticle dynamics, and Bernard van Heck, Ruben ter Meulen, Daniele Piras, and Delphine Brousse for fruitful exchanges. This research is co-funded by the allowance for Top consortia for Knowledge and Innovation (TKI’s) from the Dutch Ministry of Economic Affairs, research project {\it Scalable circuits of Majorana qubits with topological protection} (i39, SCMQ) with project number 14SCMQ02, from the Dutch Research Council (NWO), and the Microsoft Quantum initiative. CKA additionally acknowledges support from the Dutch Research Council (NWO).
\end{acknowledgments}

\section*{Data availability}
The data and analysis code that support the findings of this study will be made available at 4TU.ResearchData before final publication.\\

\section*{Author contributions}
A.B. and L.G. conceived the experiment.
Y.L. and P.K. developed and provided the nanowire materials.
A.B., M.P.V., L.J.S., J.J.W, C.K.A. and L.G. prepared the experimental setup and data acquisition tools.
A.B., L.J.S. M.P.V. and L.G. designed and fabricated the device.
A.B. and L.G. performed the measurements and analyzed the data, with continuous feedback from L.J.S., M.P.V., J.J.W, L.P.K. and C.K.A.
A.B. and L.G. wrote the manuscript with feedback from all authors.

\footnotetext[3]{See Supplemental Material at [URL], which contains further details about device fabrication, the experimental setup, device tune-up, ratios of added loss rates, and magnetic field dependence of qubit frequencies. It includes Refs. \cite{Beenakker1991, Spanton2017, Zuo2017, Heinsoo2018, Kringhoj2018, Hart2019, Bargerbos2020, Kringhoj2020}.}

\bibliography{ms.bib}

\end{document}


\myexternaldocument{ms}
\beginsupplement

\title{Supplementary information for ``Mitigation of quasiparticle loss in superconducting qubits by phonon scattering''}

\author{Arno Bargerbos}
 \email{a.bargerbos@tudelft.nl}
\affiliation{QuTech and Kavli Institute of Nanoscience, Delft University of Technology, 2628 CJ Delft, The Netherlands}
\author{Lukas Johannes Splitthoff}
\affiliation{QuTech and Kavli Institute of Nanoscience, Delft University of Technology, 2628 CJ Delft, The Netherlands}
\author{Marta Pita-Vidal}
\affiliation{QuTech and Kavli Institute of Nanoscience, Delft University of Technology, 2628 CJ Delft, The Netherlands}
\author{Jaap J. Wesdorp}
\affiliation{QuTech and Kavli Institute of Nanoscience, Delft University of Technology, 2628 CJ Delft, The Netherlands}
\author{Yu Liu}
\affiliation{Center for Quantum Devices, Niels Bohr Institute, University of Copenhagen, 2100 Copenhagen, Denmark\looseness=-1}
\author{Peter Krogstrup}
\affiliation{Niels Bohr Institute, University of Copenhagen, 2100 Copenhagen, Denmark\looseness=-1}
\author{Leo P. Kouwenhoven}
\affiliation{QuTech and Kavli Institute of Nanoscience, Delft University of Technology, 2628 CJ Delft, The Netherlands}
\author{Christian Kraglund Andersen}
\affiliation{QuTech and Kavli Institute of Nanoscience, Delft University of Technology, 2628 CJ Delft, The Netherlands}
\author{Lukas Gr\"unhaupt}
 \altaffiliation[Now at: ]{Physikalisch-Technische Bundesanstalt, 38116 Braunschweig, Germany}
\email{lukas.gruenhaupt@ptb.de}
\affiliation{QuTech and Kavli Institute of Nanoscience, Delft University of Technology, 2628 CJ Delft, The Netherlands}

\date{\today}

\maketitle

\tableofcontents

\vspace{2 cm}

\newpage

\section{Device and experimental setup}

\subsection{Device overview}
\begin{figure}[h!]
    \center
    \includegraphics[scale=1]{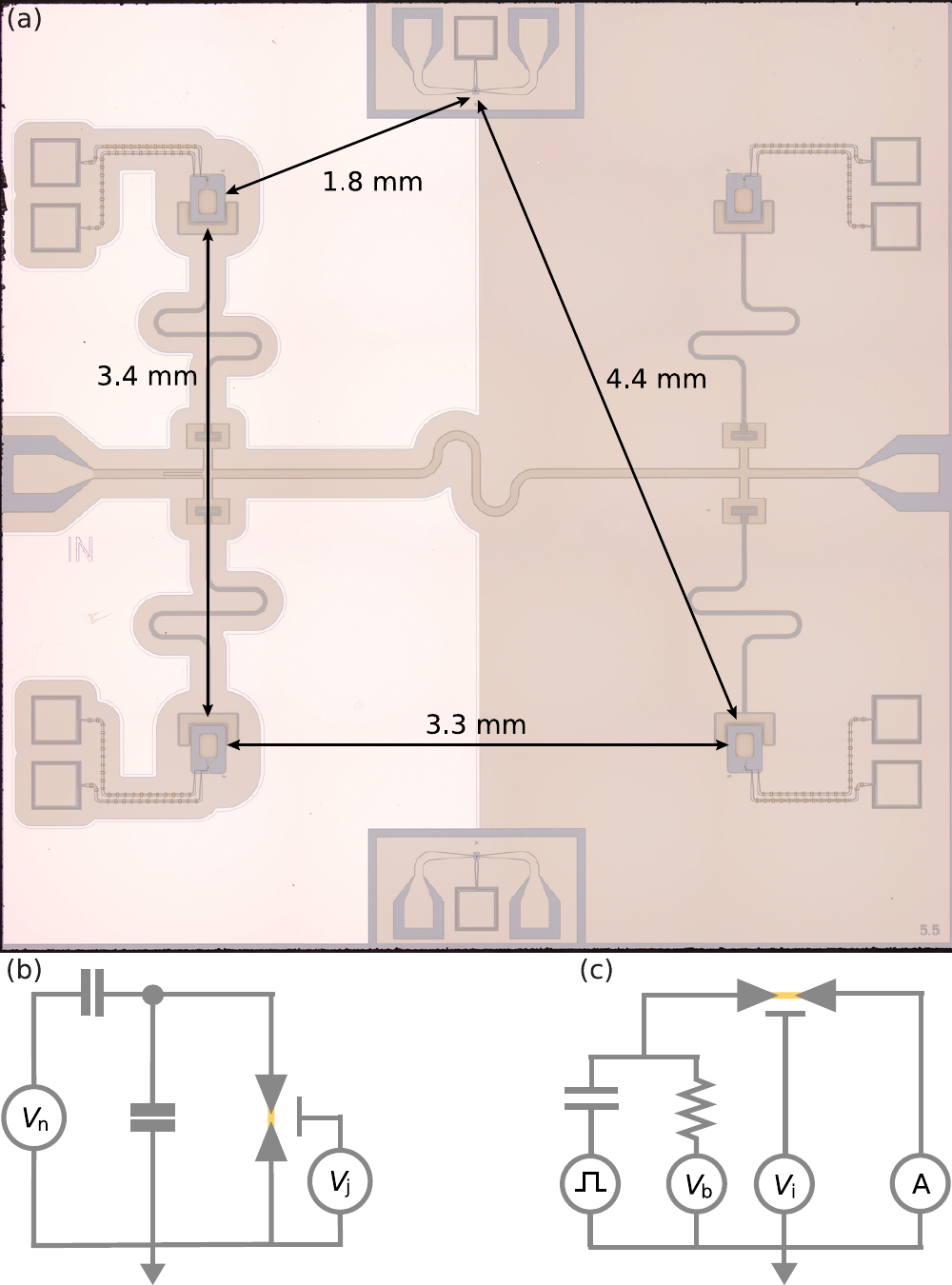}
    \caption{{\bf Extended device overview.} (a) Optical micrograph of the device before wire-bonding. The composite image is stitched together from 36 separate images. For each image the contrast is adjusted to that of its neighboring images, but no further processing is performed. (b) Circuit diagram of the nanowire transmon qubit coupled to junction and island gate electrodes. (c) Circuit diagram of the nanowire phonon injector coupled to source and drain leads as well as an additional gate electrode.}
    \label{fig:circuits}
\end{figure}

An extended optical image of the full $6 \times 6~\rm{mm^2}$ chip is shown in Fig.~\ref{fig:circuits}(a). As discussed in the main text, it contains four grounded nanowire transmon qubits coupled to individual coplanar waveguide resonators. The coplanar waveguide resonators are coupled to a common feedline, which contains a capacitor at its input port to improve the directionality of the signal \cite{Heinsoo2018}. The nanowire transmons each have two gate electrodes; the first is used to control the Josephson potential with voltage $V_{\rm j}$, the second to control the offset-charge on the island with voltage $V_{\rm n}$ [Fig.~\ref{fig:circuits}(b)]. 

At the top and bottom of the chip there are two additional nanowire junctions that are galvanically isolated from the rest of the ground plane [Fig.~\ref{fig:circuits}(c)]. The junctions are connected to two leads: the source and drain. The source is connected to a pulsed and constant voltage sources via a bias tee, while the drain is connected to a current measurement module. An additional gate electrode is present to tune the out-of-gap resistance of the junction via the field effect.

\subsection{Nanofabrication details}
The device fabrication is done using standard nanofabrication techniques. The substrate is \SI{525}{\micro\meter} thick high-resistivity silicon, on top of which a \SI{20}{nm} thick NbTiN film is sputtered. From this base layer the circuit elements are patterned using an electron-beam lithography mask and $\rm{SF_6}$/$\rm{O_2}$ reactive ion etching. In this step we also remove the NbTiN in the places where the Al phonon traps are to be placed. Immediately before deposition of the \SI{200}{nm} thick Al by electron beam evaporation, we remove surface oxide on the substrate by a \SI{30}{\second} dip in 20:1 diluted buffered oxide. The Al layer is then patterned by lift-off. After this \SI{30}{nm} of $\rm{Si_3N_4}$ dielectric is deposited to form the insulation of the junction gate electrodes using plasma enhanced chemical vapor deposition, which is patterned by wet etching with buffered oxide etch using negative electron-beam lithography. The nanowires are then deterministically placed on top of the dielectric using a nanomanipulator and an optical microscope. For this we use an approximately \SI{10}{\micro m}-long vapour-liquid-solid (VLS) hexagonal InAs nanowire with a diameter of \SI{110}{nm} and a \SI{6}{nm}-thick epitaxial Al shell covering two facets \cite{Krogstrup2015}. After placement, a \SI{100}{nm} section of the aluminium shell is selectively removed by wet etching with MF-321 developer to create the Josephson junctions. After the junction etch the nanowires are contacted by an argon milling step followed by the deposition of \SI{150}{nm} thick sputtered NbTiN. Finally, the chip is diced into $6 \times 6~\rm{mm^2}$, glued onto a solid copper block with silver epoxy, and connected to a custom-made printed circuit board using aluminium wirebonds. 

\subsection{Cryogenic and room temperature measurement setup}

\begin{figure}[h!]
    \center
    \includegraphics[scale=0.6]{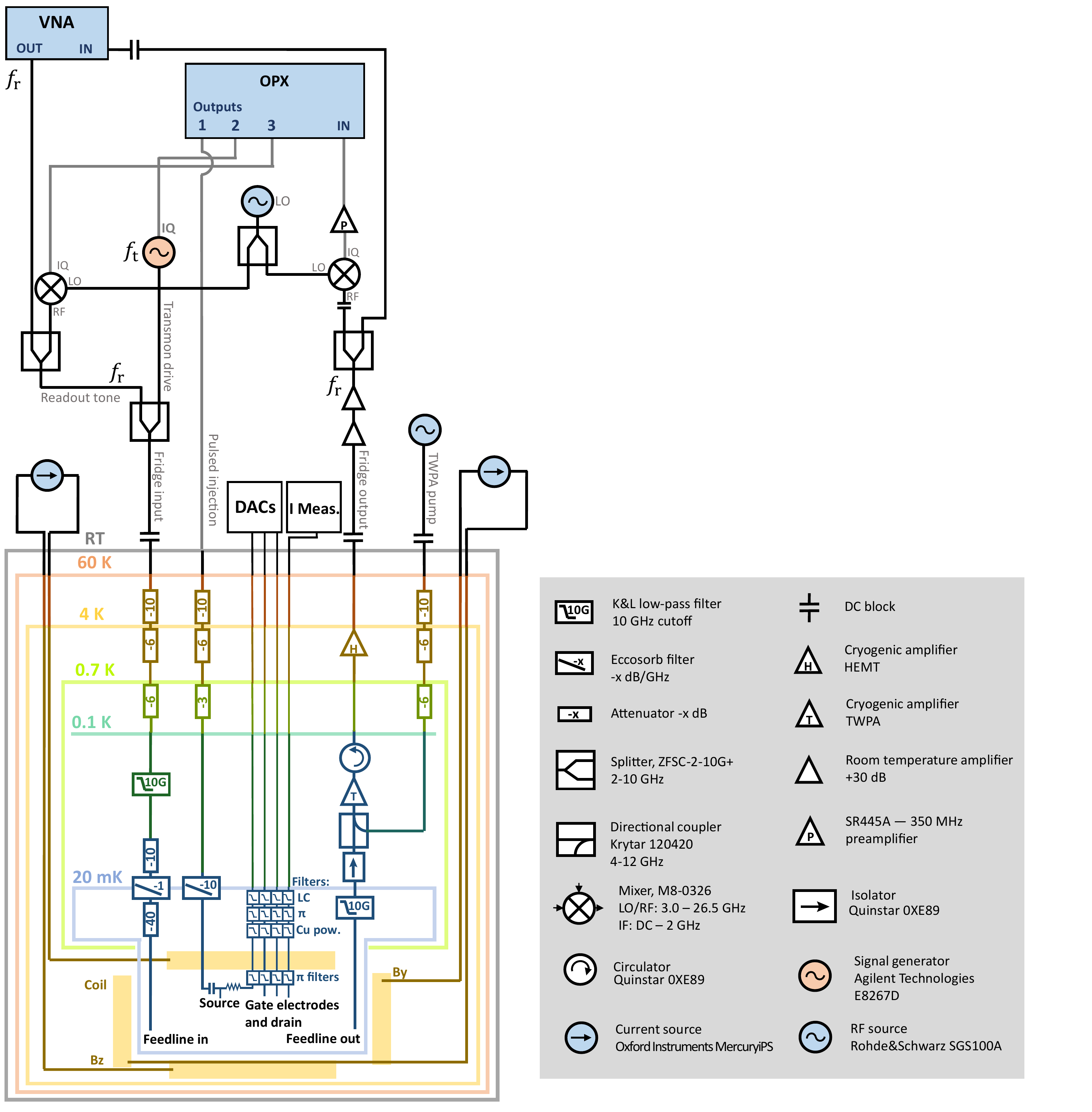}
    \caption{ Measurement setup at cryogenic and room temperatures.}
    \label{fig:cryogenic_setup}
\end{figure}

The device is measured in a commercial dilution refrigerator with a base temperature of \SI{20}{mK}. Shown in Fig.~\ref{fig:cryogenic_setup}, the setup contains an input RF line, an output RF line, an extra RF line for the pulsed injection, and multiple DC lines for voltage biasing, current sensing, and the tuning of gate voltages. Digital-to-analog (DAC) voltage sources developed in-house are connected to the DC gate electrode lines, which are filtered at base temperature with multiple low-pass filters. The injector junction drain lead contains the same filtering and is connected to a current measurement module also developed in-house. It is further connected to a Keithley 2400 multimeter and a Stanford Research Instruments SR830 Lock in amplifier (not shown). The input and pulse RF lines contain attenuators and filters at different temperature stages, as indicated. The output RF line contains a traveling wave parametric amplifier (TWPA) acquired from the MIT Lincoln Laboratory at the \SI{20}{mK} temperature stage, a Low Noise Factory high-electron-mobility transistor (HEMT) amplifier at the \SI{4}{K} stage, and two additional Narda-MITEQ low noise amplifiers at room temperature. A three-axis vector magnet (x-axis not shown) is thermally anchored to the \SI{4}{K} temperature stage, with the device under study mounted at its center. The three magnet coils are  controlled with MercuryiPS current sources. At room temperature, a vector network analyzer (VNA) is connected to the input and output RF lines for spectroscopy at frequency $f_{\rm r}$. On the input line, this signal is then combined with the IQ-modulated transmon drive tone at frequency $f_{\rm t}$. A separate IQ-modulated tone at $f_{\rm r}$, only used for time-domain measurements, is also combined onto this line. The pulsed injection signal is directly sent into the fridge at baseband frequencies. For time-domain measurements the output signal is additionally split off into a separate branch measured with a commercial quantum system measurement platform. 

\clearpage 

\section{Basic characterization}
\subsection{Qubits}
In this section we discuss the characterization of the four qubits at the setpoints used in the experiments. In practice, the choice for the setpoints is governed by several factors: the frequency dependence of the qubit lifetimes through Purcell and dielectric losses, the suppression of offset-charge sensitivity to allow for parity-insensitive qubit manipulation, and the need for inter-qubit frequency detuning to allow for individual control of the qubits through a single feedline. Under these constraints, we empirically find the optimal range of qubit frequencies to be in the range \SIrange{3.7}{4.0}{GHz}. The resulting readout resonator frequencies $f_{\rm r}$, qubit frequencies $f_{01}$, and average qubit lifetimes $T_1$ in the absence of phonon injection are listed in Table \ref{tab:qubpars}.

However, in order to compare the effect of phonon injection on each qubit on equal footing, we further need to consider that the interaction between the phonon-induced quasiparticles and the qubits depends on the values of the qubit parameters. Specifically, for transmon qubits based on superconductor-insulator-superconductor (SIS) junctions with a Josephson energy $E_{\rm J}$ significantly larger than the charging energy $E_{\rm c}$, the added qubit loss due to quasiparticles is given by \cite{Glazman2021}
\begin{equation}
\Gamma_{10} = \frac{16 E_{\rm J}}{\hbar \pi}\sqrt{\frac{E_{\rm c}}{8E_{\rm{J}}}} \sqrt{\frac{\Delta}{2 hf_{01}}} x_{\rm{qp}} = D x_{\rm{qp}}
\label{eq:qpprate}
\end{equation}
where $f_{01}$ is the qubit frequency and $x_{\rm{qp}}$ the density of quasiparticles. While the validity of this equation for the semiconducting Josephson junctions used in this work has not been investigated, we choose our qubit setpoints in accordance with this equation in an attempt to ensure comparable sensitivity to quasiparticles (see the estimated values of $D$ in Tab.~\ref{tab:qubpars}).

\begin{figure}[h!]
    \center
    \includegraphics[scale=1]{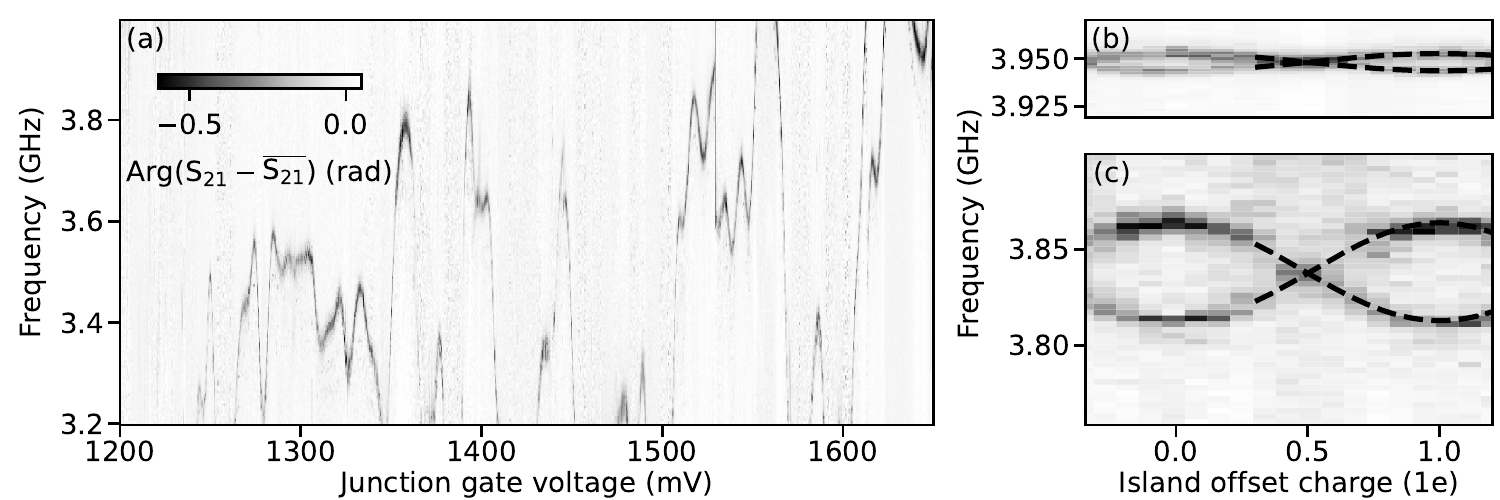}
    \caption{Characterization of the nanowire transmon qubits. (a) Representative junction gate dependence of the $\rm{Near}_{Al}$ qubit near its setpoint at \SI{1659}{mV} used in the main text. (b-c) Offset charge dependence of the $f_{01}$ (top) and $f_{02}/2$ transition frequencies of the $\rm{Near}_{Al}$ qubit as capacitively tuned with the island gate electrode, measured at the setpoint used in the main text. Partially overlaid are fits with Eqs.~\eqref{eq:transmon_ham} and \eqref{eq:feigelman}, used to determine the parameters listed in Tab.~\ref{tab:qubpars}.}
    \label{fig:qubittuneup}
\end{figure}

The tuning of the qubits is performed in-situ via the field effect, making use of the bottom gate electrode located below the qubit's Josephson junction \cite{deLange2015, Larsen2015}. As shown in Fig.~\ref{fig:qubittuneup}(a), this results in mesoscopic fluctuations of the qubit frequency as a function of gate voltage, allowing for fine tuned control. When possible, we choose to place qubits at a local maximum or minimum in gate voltage, reducing sensitivity to charge noise \cite{Luthi2018}. Next we characterize the Hamiltonian parameters of the qubits in more detail by fitting the measured qubit frequency. This is done by numerically diagonalizing the transmon Hamiltonian
\begin{equation}
H = -4E_{\rm c}\partial_\phi^2 + V(\phi)
\label{eq:transmon_ham}
\end{equation} 
where $V(\phi)$ is the Josephson potential of the junction, dependent on the superconducting phase difference across the junction $\phi$. For SIS junctions it takes on the simple form $V(\phi)  = E_{\rm J}\left[1-\cos(\phi)\right]$, allowing one to uniquely determine the transmon's Hamiltonian parameters through a measurement of the qubit frequency $f_{01}$ and its anharmonicity $ f_{21} - f_{01}$. However, for the  semiconducting weak link Josephson junctions employed in the experiment the situation is more complex. Here the Josephson potential is governed by a limited number of possibly highly transparent Andreev bound states, requiring a more complex potential strongly dependent on the microscopics of the junction \cite{Kringhoj2018}. In the limit where the length of the junction is much shorter than the coherence length, one often employs a potential of the form
\begin{equation}
V(\phi) = -\sum_{i=1}^N \Delta_i \sqrt{1-T_i \sin^2\left(\phi/2\right)}.
\end{equation}

This equation describes $N$ non-interacting channels with normal state transmission $T_i$ and superconducting gap $\Delta_i$ \cite{Beenakker1991}. For the case of a a single channel this type of expression has been shown to lead to good agreement with experiments, but when $N>1$ it becomes highly non-trivial to uniquely determine the junction parameters from the transmon spectra alone \cite{Spanton2017, Hart2019}. In order to circumvent this problem we operate the qubits near the pinch-off voltage of the junctions, where only a single level strongly contributes. Nevertheless, this leads to an additional complication: near pinch-off, the behavior of the junction is typically governed by accidental quantum dots, resulting in near-unity transparencies due to resonant tunneling \cite{Bargerbos2020, Kringhoj2020}. In this case the charge dispersion of the transmon qubit can be drastically reduced beyond what is expected from its effective Josephson energy due to non-adiabatic phase dynamics. This effect is also encountered in our experiment, and has to be included in order to accurately fit the data. For this we extend the Josephson potential to include the presence of the higher lying Andreev bound states:
\begin{equation}
V(\phi)    =
\widetilde{\Delta} \begin{pmatrix}
    \cos{\frac{\phi}{2}} & \sqrt{1-T}\sin{\frac{\phi}{2}} \\[4pt]
    \sqrt{1-T}\sin{\frac{\phi}{2}}  & -\cos{\frac{\phi}{2}} 
\end{pmatrix}
\label{eq:feigelman}
\end{equation}
Here $\widetilde{\Delta}$ is an effective gap energy of the Andreev bound states, potentially reduced below that of the superconducting leads $\Delta_{\rm nw}$ by confinement and charging effects \cite{Bargerbos2020, Kringhoj2020}.

Having established the model for the nanowire transmon qubits, we now characterize the qubit parameters by measuring the remaining offset-charge dependence of the $f_{01}$ and the two-photon $f_{02}/2$ transition frequencies for each of the qubits. We then diagonalize the Hamiltonian of \eqref{eq:transmon_ham} using the potential of \eqref{eq:feigelman} and fit this to the measurements, resulting in a unique set of qubit parameters. These parameters are also tabulated in Table \ref{tab:qubpars}. We find that the qubits have comparable charging energies, close to the targeted value of $\SI{400}{MHz}$. Furthermore, their effective Josephson energies $E_{\rm J}^{\rm eff} = \widetilde{\Delta} T/4$ are again similar, resulting in comparable quasiparticle proportionally constants $D$. Although the validity of Eq.~\eqref{eq:qpprate} for weak link Josephson junctions is not strictly established, these estimates indicate that the chosen qubit setpoints do not bias the experiment towards positive effects of the phonon traps. Based on the obtained values of $D$ and the loss rates reported in main text Fig.~2(d), we further estimate that the phonon injection creates non-equilibrium quasiparticle densities $x_{\rm qp}$ of up to $5 \times 10^{-4}$ at the $\rm{Near}_{\rm NbTiN}$ qubit.

\begin{table}[t!]
\centering
\begin{tabular}{|l|l|l|l|l|l|l|}
\hline
Qubit & $f_{r}$~(GHz)& $f_{01}$~(GHz) & $T_1$~($ \SI{}{\micro s}$) & $E_{\rm J}^{\rm eff}/h$~(GHz) & $E_{\rm J}^{\rm eff}/E_c$ & $D~(\rm ns^{-1})$ \\ \hline
$\mathrm{Near}_{\rm{Al}}$  & 7.028 & 3.948 & 3.4 & 5.26 & 13.1 & 7.5    \\ \hline
$\mathrm{Near}_{\rm{NbTiN}}$  & 6.873 & 3.864 & 2.6 & 5.07 & 12.8 & 7.4   \\ \hline
$\mathrm{Far}_{\rm{NbTiN}}$  & 6.803 & 3.784 & 4.0 & 5.15 & 13.8 & 7.3   \\ \hline
$\mathrm{Far}_{\rm{Al}}$  & 6.762 & 3.892 & 4.5 & 4.92 & 11.7 & 7.5     \\ \hline       
\end{tabular}
\caption{Extracted qubit parameters at zero applied magnetic field. Estimated from the qubit's charge dispersion through Eqs.~\eqref{eq:qpprate} and \eqref{eq:feigelman}.}
\label{tab:qubpars}
\end{table}

\subsection{Injector}
\begin{figure}[h!]
    \center
    \includegraphics[scale=1]{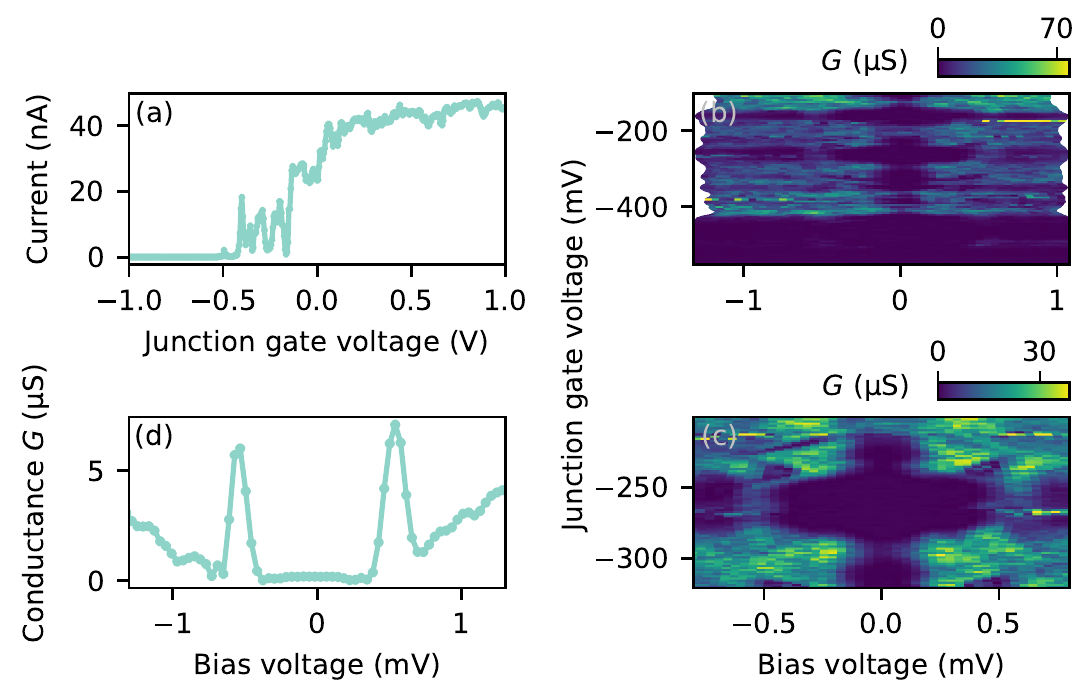}
    \caption{Characterization of the injector Josephson junction. (a) Pinchoff curve of the measured current at $V_{\rm bias}=\SI{1}{mV}$. (b-c) 2D maps of differential conductance $G$ as a function of the junction gate voltage and the bias voltage, where (c) shows a higher resolution zoom-in of the region used in the experiment. (d) Differential conductance versus bias voltage at a junction gate voltage of \SI{-251}{mV}, the setpoint used in the experiment. Note that this panel is not a linecut of panel (c); it is measured simultaneously with main text Fig.~2(a,c), and its position is slightly shifted with respect to the other panels due to hysteresis.}
    \label{fig:injtuneup}
\end{figure}

We now discuss the setpoint chosen for the phonon injection junction. To start, we measure the so called  `pinchoff curve' of the junction by applying a constant bias voltage of $V_{\rm bias} = \SI{1}{mV}$, several times beyond the expected superconducting gap of the nanowire. We then measure the current while varying the gate voltage applied to the junction [Fig.~\ref{fig:injtuneup}(a)]. For negative voltages, no current flows, until around \SI{-0.5}{V} where we observe an oscillatory onset of current. For larger gate voltages the current grows, eventually saturating and reaching several tens of nano-amperes. The oscillatory behaviour near pinch-off is once-more indicative of accidental quantum dots forming inside the junction region, similar to that seen in Fig.~\ref{fig:qubittuneup}(a). 

Further insight into the behavior of the junction is obtained by measuring the bias voltage dependent differential conductance $G$ using lock-in amplification. As shown in Fig.~\ref{fig:injtuneup}(b-c), around zero bias voltage the junction portrays a region of strongly reduced conductance, indicative of a transport gap. The width of this region as a function of junction gate voltage is set by the superconducting gap of the nanowire $\Delta_{\rm nw}$, modified by an interplay of Coulomb effects and multiple Andreev reflection. For the setpoint used in the experiments we fix the junction gate voltage in the regime where the size of the transport gap is maximal, resulting in an estimated $\Delta_{\rm nw}=\SI{270}{\micro eV}$ [Fig.~\ref{fig:injtuneup}(d)]. We note that the size of this gap could include residual effects of Coulomb blockade, however the value extracted is comparable to results found on nanowires from the same growth batch \cite{Splitthoff2022}.

\clearpage
\section{Extended data}
\subsection{Ratios of loss rates}
\begin{figure}[h!]
    \center
    \includegraphics[scale=1]{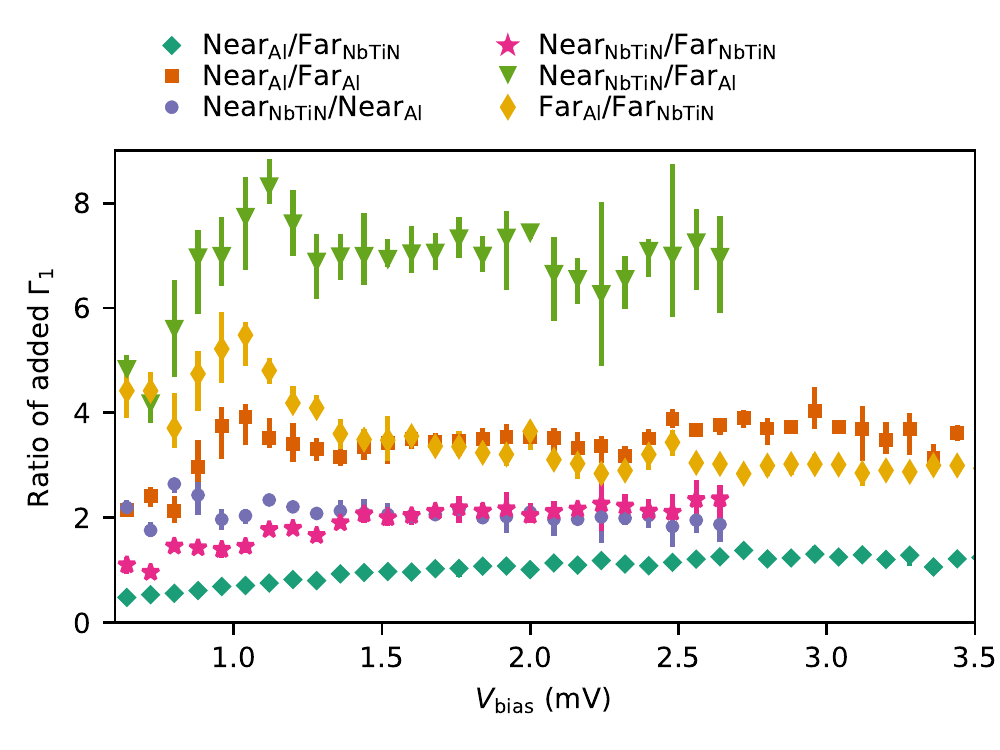}
    \caption{Ratio of added loss rates as measured for constant injection voltages at zero applied magnetic field [c.f. Fig.~2(d)]. The colors and symbols used in this figure do not match those used to identify the qubits throughout the text. }
    \label{fig:ratios}
\end{figure}

Fig.~\ref{fig:ratios} shows different ratios of added loss rate in the presence of a constant bias voltage across the injector junction. All six scenario's are compared. As discussed in the main text, this shows that the largest difference occurs between the $\mathrm{Near}_{\rm{NbTiN}}$ and the $\mathrm{Far}_{\rm{Al}}$ qubits, reaching up to 8 times more added loss for the $\mathrm{Near}_{\rm{NbTiN}}$ qubit, although this is potentially due to a combination of distance from the injector as well as the presence of the phonon traps. On equal footing, the $\mathrm{Near}_{\rm{Al}}$ qubit has up to 2.5 times smaller added loss rate than the $\mathrm{Near}_{\rm{NbTiN}}$ qubit, and the  $\mathrm{Far}_{\rm{Al}}$ has up to 5 times smaller loss than the $\mathrm{Far}_{\rm{NbTiN}}$ suggesting that an increased area of the trapping region increases the efficacy. Furthermore, the $\mathrm{Far}_{\rm{NbTiN}}$ qubit performs up to 2 times better than the  $\mathrm{Near}_{\rm{NbTiN}}$. As both qubits are not directly next to the Al phonon traps, this supports previous findings that qubit errors are strongest close to the position where the phonons originate \cite{McEwen2021, Martinis2021, Wilen2021}. Finally, we note that the ratio of added loss rates is a function of the bias voltage, the exact cause of which remains to be investigated.

\subsection{Magnetic field dependence}
\begin{figure}[h!]
    \center
    \includegraphics[scale=1]{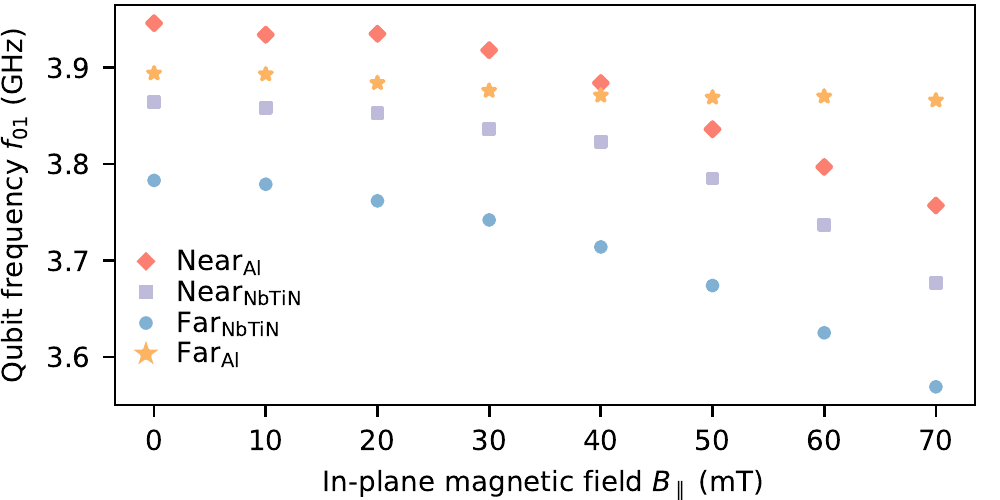}
    \caption{Magnetic field dependence of nanowire transmon qubit frequencies $f_{01}$. }
    \label{fig:sfield}
\end{figure}

As discussed in the main text, we make use of the inherent magnetic field compatibility of the nanowire transmon qubits and their readout circuit to investigate the effect of the Al phonon traps turning normal. In previous works, nanowire-based superconducting qubits have been shown to be operable up to parallel fields in excess of \SI{1}{T} \cite{Luthi2018, PitaVidal2020, Uilhoorn2021}. In Fig.~\ref{fig:sfield} we show the evolution of the four nanowire transmon qubits used in this work as a function of the magnetic field. Up to \SI{70}{mT}, only modest changes in qubit frequency of less than 5\% are found, leading to a negligible change in the quasiparticle proportionality constant $D$ of Eq.~\eqref{eq:qpprate}. This shows that the absence of changes in the added loss rates versus magnetic field can indeed be attributed to the phonon traps having a constant effect at all investigated fields, rather than to a field dependence canceled out by changes in qubit sensitivity. Furthermore, we note that the qubits do not all have the same field evolution; in particular the $\mathrm{Far}_{\rm{Al}}$ qubit appears more resilient over the range of fields explored. We hypothesize this occurs because the magnetic field evolution of the nanowires is governed not only by the reduction of the superconducting gap in the Al shell, but also by orbital effects that depend on the spatial profile of the occupied transport channels inside the nanowire \cite{Zuo2017}. 

\bibliography{ms.bib}